\documentclass[preprint]{aastex}

\usepackage{color,graphicx}
\usepackage{amssymb,amsmath,eurosym,psfrag,abstract}
\usepackage{mathrsfs}
\usepackage{enumerate}
\usepackage{natbib}

\usepackage[pdftex,plainpages=false]{hyperref} 
\usepackage[all]{hypcap}

\newcommand{\figref}[1]{Figure \ref{#1}}
         
\newcommand{\Frac}[2]{\frac{\raisebox{-0.5ex}{\ensuremath{#1}}}{#2}}

\newcommand{\vect}[1]{\mathbf{#1}}

\newcommand{\Matrix}[1]{\mathbb{#1}}

\newcommand{\fnc}[1]{\textsf{#1}} 
\newcommand{\unit}[1]{\ensuremath{\, \mathrm{#1}}}

\def\bigvar#1#2{{\hbox{$\left#1\vbox to #2{}\right. \n@space$}}}
\def\n@space{\nulldelimiterspace=0pt \m@th}
\def\m@th{\mathsurround=0pt }

\author{Lucas Tarr \& Dana Longcope \& Margaret Millhouse}
\affil{Department of Physics, Montana State University, Bozeman, Montana 59717}
\title{Calculating Separate Magnetic Free Energy Estimates for Active Regions Producing Multiple Flares: NOAA AR11158}
\date{Draft: \today}

\begin{document}
\maketitle
\begin{abstract}
  It is well known that photospheric flux emergence is an important process for stressing coronal fields and storing magnetic free energy, which may then be released during a flare.  The \emph{Helioseismic and Magnetic Imager} (HMI) onboard the \emph{Solar Dynamics Observatory} (SDO) captured the entire emergence of NOAA AR 11158.  This region emerged as two distinct bipoles, possibly connected underneath the photosphere, yet characterized by different photospheric field evolutions and fluxes.  The combined active region complex produced 15 GOES C--class, 2 M--class, and the X2.2 Valentine's Day Flare during the four days after initial emergence on February 12th, 2011.  The M and X class flares are of particular interest because they are nonhomologous, involving different subregions of the active region.  We use a Magnetic Charge Topology together with the Minimum Current Corona model of the coronal field to model field evolution of the complex.  Combining this with observations of flare ribbons in the 1600\AA\ channel of the \emph{Atmospheric Imaging Assembly} (AIA) onboard SDO, we propose a minimization algorithm for estimating the amount of reconnected flux and resulting drop in magnetic free energy during a flare.  For the M6.6, M2.2, and X2.2 flares, we find a flux exchange of $4.2\times 10^{20}\unit{Mx},\ 2.0 \times 10^{20}\unit{Mx} , \hbox{and } 21.0 \times 10^{20}\unit{Mx}$, respectively, resulting in free energy drops of $3.89\times 10^{30}\unit{ergs}, 2.62\times 10^{30}\unit{ergs}, \hbox{and } 1.68\times 10^{32}\unit{ergs}$.

\end{abstract}

\section{\label{sec:intro}Introduction} 

Solar flares are the most extravagant examples of rapid energy release in the solar system, with the largest releasing around $10^{32}\unit{ergs}$ on a timescale of hours \citep{Benz:2008}.  This energy, imparted to the plasma confined along coronal magnetic loops of active regions, is distributed between kinetic, thermal, and radiative process in some way that may vary from flare to flare.  While the ultimate source of this energy is likely stresses introduced by convective motion of the plasma at and below the photosphere, we believe the direct source is the conversion of free magnetic energy: magnetic energy in excess of the active region's potential magnetic field energy.

As has been clear for many decades, active regions consist of bundles of flux tubes, concentrated prior to their emergence through the photosphere \citep{Zwaan:1978}.  The free energy builds up as the flux tubes forming an active region are stressed at the photospheric boundary, where plasma forces dominate field evolution (plasma $\beta\equiv 8\pi p/B^2 > 1$, with $p$ the gas pressure).  Moving outward from the solar surface into the corona, the plasma pressure rapidly diminishes and magnetic forces dominate, until a third regime is reached where plasma forces once again dominate.  As noted by \citet{Gary:2001}, even within an active region, the high $\beta$ portion of the upper corona may occur as low as $200\unit{Mm}$ above the solar surface.  We are primarily concerned with lower laying loops and magnetic domains and so will assume a low $\beta$ regime.  Barring any reconfiguration of the coronal field, the active region's magnetic domains are pushed into a highly nonpotential state.  Relaxation towards a potential field configuration through magnetic reconnection then allows for the conversion of magnetic free energy into kinetic and thermal energy through, e.g., field line shortening, shock formation, electron acceleration, or (possibly) ion acceleration \citep{Longcope:2009a, Guidoni:2010, Fletcher:2008, Hudson:2012}.

The number of quantitative estimates of this energy buildup using observations has recently increased, but results remain varied.  Nonlinear force--free models \citep[NLFF:][]{Sun:2012, Gilchrist:2012} have received much attention during the past decade, strongly driven by both increases in computing power and the arrival of vector magnetograms from space--based telescopes onboard Hinode (SOT/SP) and SDO (HMI).  While these models are a promising avenue of research, they come with their own set of problems, as discussed in \citet{DeRosa:2009}.  The lower boundary conditions are, in general, incompatible with the force--free assumption \citep{Metcalf:1995}.  Several methods exist to overcome this difficulty \citep{Wheatland:2009}, leading to different energy estimates for a single vector magnetogram, even when using a single extrapolation code\citep{DeRosa:2009}.  

A further problem is that the models amount to a series of independent fields at consecutive timesteps.  At each time, a new NLFF field is generated from the boundary data, uninformed by the solution from the previous timestep.  Contrasting with this are flux transport and magneto--frictional models, which do include a memory \citep{Yang:1986, MGvB:2011}.  These methods primarily focus on the global coronal response to active region emergence, destabilization, and eruption as opposed to the detailed analysis of processes within an active region, which is the topic of this investigation \citep{Yeates:2008}.  One reason for this is that the coronal portion of these models evolve the large--scale mean field using an induction equation with an effective magnetic diffusivity \citep{Ballegooijen:2000}, so that the formation of fine--scale current sheets is beneath their resolution.  Most dynamical simulations without magnetic diffusion show a tendency toward fine layers \citep{VanBallegooijen:1985}.

We describe the coronal field using the Magnetic Charge Topology (MCT) model \citep{Baum:1980,Longcope:2005} at each time.  The system is described by a set of unipolar regions.  The distribution of magnetic flux between each pair of oppositely signed regions defines the system's connectivity.  As the active region evolves its connectivity will generally change.  To relate each time with the next, we employ the Minimum Current Corona model \citep[MCC:][]{Longcope:1996,Longcope:2001}.  By itself, MCT describes only potential fields, which contain no current.  The MCC introduces currents, and the resulting energetics, into the MCT model by asserting that the coronal field move through a series of Flux Constrained Equilibria (FCE).  In that case, the connectivity of the real field will be different from the potential field's connectivity.

One shortcoming of the MCC method as currently used is its inability to account for violation of these flux constraints, which are the topological manifestations of reconnection and the resulting energy release.  Previous studies \citep{Tarr:2012, Kazachenko:2012, Kazachenko:2010, Kazachenko:2009} have therefore only reported the total free energy difference between the MCC and a potential field configurations.  Our goal here is to relax those flux constraints at any timestep, while also allowing the system to continue evolving thereafter.  In this way, we may model multiple reconnection events for a single active region.

We present here a method for identifying the magnetic domains activated in successive flares based on observations of flare ribbons in the AIA 1600\AA\ channel.  This allows us to separately calculate the free energy available to each successive, nonhomologous flare.  If we further assume that all magnetic flux topologically capable of transferring during a reconnection event does transfer, then we may also estimate the actual energy release during a flare.

In the following sections, we will describe the data used (\S \ref{sec:data}), our methods for modeling the photospheric and coronal fields (\S\ref{sec:model}), how one may estimate the MCC free energy based on those models (\S\ref{sec:energy}), and the use of observations of flare ribbons to determine those domains activated in successive flares (\S\ref{sec:ribbons}).  We will conclude with a discussion of the results of our analysis (\S\ref{sec:disc}).

\section{\label{sec:data}Data}
To construct the MCC model of magnetic field evolution we use a series of 250 line--of--sight (LOS) magnetograms taken by the Helioseismic and Magnetic Imager \citep[HMI:][]{Schou:2012,Scherrer:2012,Wachter:2012} onboard the Solar Dynamics Observatory (SDO).  The data are at a 24 minute cadence between Feb.~11 2011 08:10:12 UT and Feb.~15 11:46:12 UT, and are taken from the JSOC \textsf{hmi.M\_720s} (level $1.5$) data series.  The region considered, NOAA AR11158, produced the first GOES X--Class flare of solar cycle 24, and has therefore already been analyzed in a variety of ways by numerous authors \citep[see][and references therein]{Petrie:2012}.

In addition we have used images of flare ribbons observed with Atmospheric Imaging Assembly (SDO/AIA) in the 1600\AA\ channel \citep{Lemen:2011}.  We obtained three sets of 1600\AA\ images via the SSW cutout service maintained by Lockheed Martin\footnote{\url{http://www.lmsal.com/get_aia_data/}} for $\approx 30\unit{min}$ during each flare with peak magnitude greater the M1.0: an M6.6 flare peaking at Feb 13, 17:28; M2.2 peaking at Feb 14, 17:20; and X2.2 flare peaking at Feb 15, 01:44.  All AIA data were prepared to level $1.5$ using the standard \fnc{aia\_prep} routine in SolarSoftWareIDL.  

We coalign each set of AIA 1600\AA\ images to the magnetogram closest to the peak time of each flare.  To do so, we rotate the first AIA image in a sequence to the time of the magnetogram.  It so happens that the 75G contour of HMI LOS magnetograms (after assuming a radial field and correcting for pixel foreshortening, discussed below) outlines the bright network patches in the 1600\AA\ band.  We shift the AIA image, by eye, until the contour and bright network patches align.  This could be automated by a cross--correlation between the magnetogram contour and a corresponding contour in AIA, but we have not yet implemented this procedure.  The AIA timesets are internally aligned, so we apply the same by--eye offset to each image in the sequence, after shifting each by solar rotation to the time of the chosen magnetogram.

The HMI magnetograms contain known (but as yet unmodeled) diurnal variations, due to the velocity of the spacecraft's orbit \citep{Liu:2012}.  The amplitude of the variations is around $2.5\%$ of the unsigned flux within an active region.  We are concerned with flux emergence trends over the course of days, over which time the effects of these variations should largely cancel.  We simply accept this as an additional source of error in our model.

\section{\label{sec:model}Modeling the Magnetic Evolution}

We apply the methods described in \citet{Tarr:2012} and references therein to generate the magnetic modeling for this series of events.  The analysis splits into two sections, detailed below.  First, we characterize the photospheric field by partitioning the observed magnetograms into a set of unipolar regions.  Pairs of oppositely--signed regions $\{j,k\}$ may be linked through emergence when each region's flux increases between two timesteps.\footnote{Regions may also submerge or diffuse, as P1 does after $t\approx 25\unit{hrs}$.  Algorithmically, there is no distinction between these processes.  If any of a pole's flux change between two timesteps cannot be paired with another region, it is formally paired with a source of opposite sign located at infinity.}  At each time $i$, the amount of flux change of each pairing is recorded in a photospheric--field--change matrix $\Delta^i\Matrix{S}_{j,k}$.  This set of matrices is therefore a time history of the flux with which each region emerged with every other region.

In the second part of our analysis, we develop a topological model of coronal domains immediately prior to each major flare.  Our flux emergence matrix discussed above is the real flux in each coronal domain, which we may compare to the flux in each domain in a potential field extrapolation.  The difference between the two is the nonpotentiality of each domain.  The equilibrium with minimum magnetic energy that still includes this difference in domain fluxes, called the flux constrained equilibrium (FCE), contains current sheets on each of its separators \citep{Longcope:2001}.  Our topological model determines the location of all current sheets within the active region complex, the strength of each related to the nonpotentiality of its associated domains.  Finally, this provides us with an estimate of the energy in the FCE, which is itself a lower bound on the magnetic free energy stored in the actual magnetic field.

\subsection{\label{sec:mod-phot}Modeling the Photospheric Field}
To characterize the photospheric field, we first convert each line of sight (LOS) pixel to vertical by assuming a radial field and account for reduced flux due to pixel foreshortening dividing the flux in each pixel by $\cos^2(\theta)$, where $\theta$ is the polar angle from disk center.  We smooth our data by extrapolating the vertical field to a height of $3\unit{Mm}$ as detailed in \citet{Longcope:2009}, and reduce noise by ignoring any pixels below a $75\unit{G}$ threshold.  Pixels above this threshold are partitioned using the downhill tessellation algorithm of \citet{Barnes:2005}, creating a \emph{mask}.  Each pixel is assigned an integer, and contiguous groups of like--signed pixels of the same integer compose a \emph{region}.  Pixels below our threshold belong to no region and are assigned a mask value of 0.

This tessellation scheme can generate thousands of small regions at each timestep, so adjacent regions of like polarity are merged when the saddle point value of the magnetic field between them is less than $700\unit{G}$.  Finally, we exclude any region whose total flux is less than $2\times10^{20}\unit{Mx}$.  This process is carried out at each timestep, independently.  We call the resulting set of masks a \emph{mask array}.  Regions in one timestep are associated with those in the surrounding timestep first by a bidirectional association between timesteps, and then applying the \fnc{rmv\_flick} and \fnc{rmv\_vanish} algorithms, described in detail in \citet{Tarr:2012}.  

The focus of these methods is to distinguish between regions of flux that emerge from below the photosphere at different times and in different places, and to keep track of these individual regions as they undergo shear motion on the solar surface after emergence.  To this end, after the automatic algorithms listed above have run to convergence, we manually shift the boundaries between regions to ensure we consistently track flux emergence and migration over the entire timeseries.  We allow individual regions to emerge, submerge, change shape, translate, split, and merge.  To further reduce the number of regions, we exclude regions which have non--zero flux for fewer than 5 timesteps ($\approx 100\unit{min}$).  

\begin{figure}[ht]
  \begin{center}
    \includegraphics[width=0.48\textwidth]{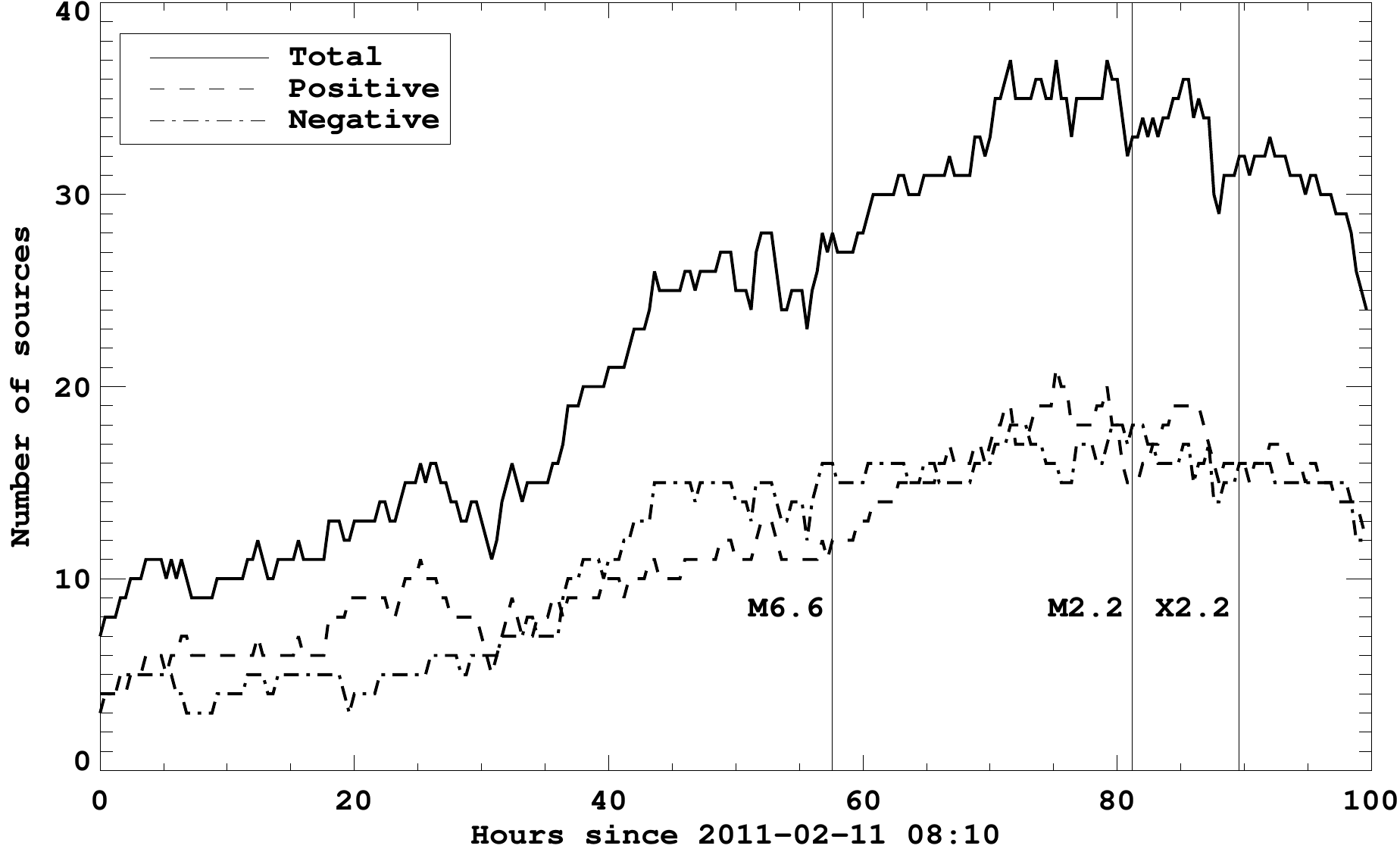}
    \caption[Sources]{\label{fig:sources} Number of distinct positive, negative, and total mask regions at each time.}
  \end{center}
\end{figure}
\figref{fig:sources} shows the number of mask regions of each polarity over the entire timeseries.  In our final analysis, we track 118 regions for $\approx 100$ hrs.  The number of distinct regions at any time varies between 10 and 35, generally growing at a steady rate as the active region complex emerges between 2011-02-11T08:10:12 and 2011-02-14T16:10:12, then dropping slightly as the fully emerged system continues to concentrate\footnote{\figref{fig:sources} shows that for AR11158, neither polarity is consistently or significantly more fragmented than the other.  This contrasts with observations of most other active regions, where the leading polarity is substantially more concentrated than the trailing.  This can show up in MCT models as the leading polarity's flux distributed amongst fewer poles than that trailing polarity.  This does not seem to be the case for AR11158, which we find curious, although it does not affect our analysis.}.

\begin{figure*}[ht]
  \begin{center}
  \capstart
  \includegraphics[width=0.8\textwidth]{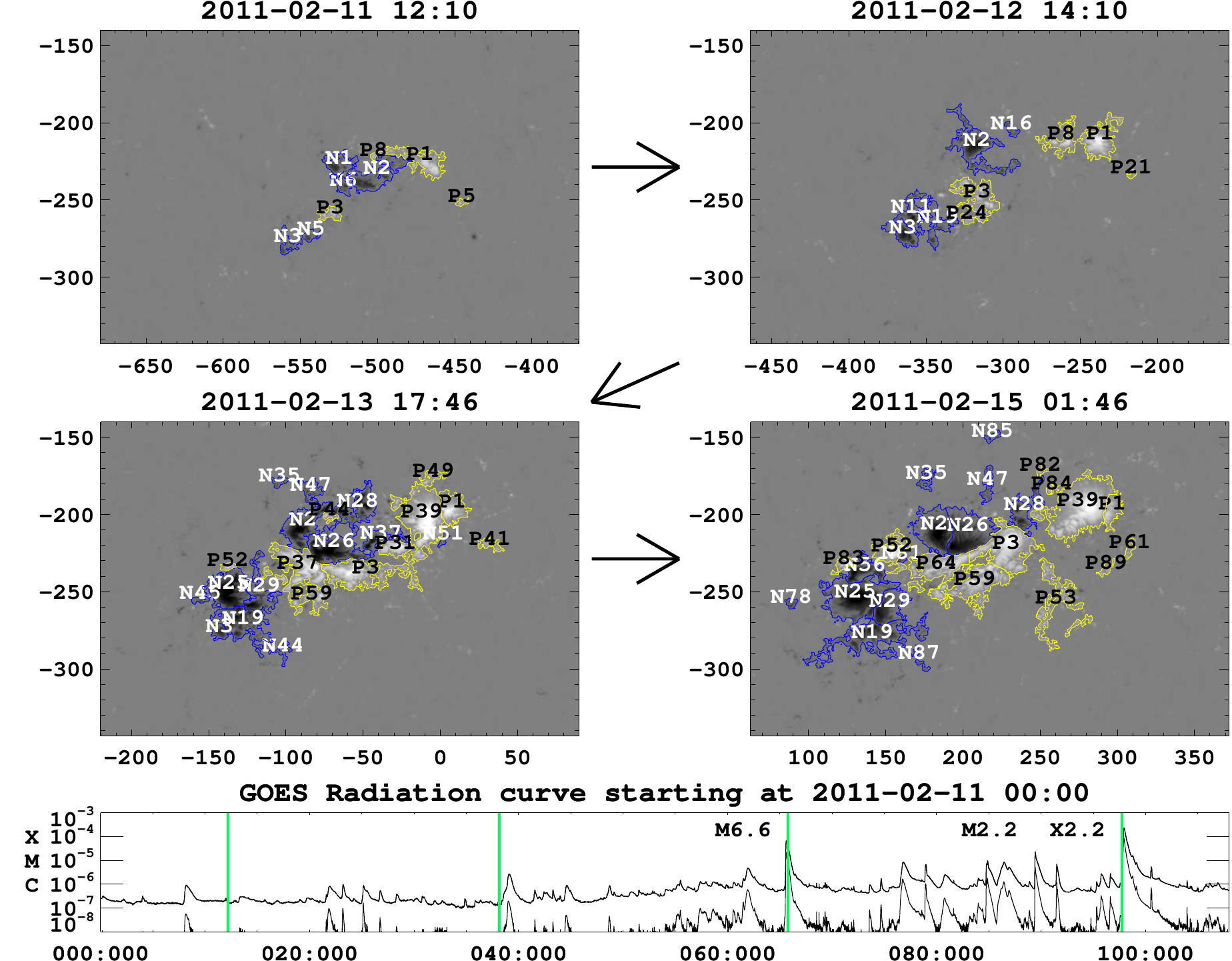}
  \caption[NOAA AR 11158]{\label{fig:partition}  \emph{Top:} HMI LOS magnetograms of NOAA AR 11158 overlaid with their respective masks.  The grayscale saturates at $\pm max(\vert B\vert )=\pm 1500.0$G, and the axes are in arcseconds from disk center.  Arrows indicate the time sequence.  \emph{Bottom:} Full--disk integrated GOES electron radiation curve.  The four vertical lines correspond to the times of the four displayed magnetograms.}
  \end{center}
\end{figure*}

Four examples of the LOS magnetogram overlaid with masks are shown in \figref{fig:partition}.  The top four images are each a single frame from an animation of the full 250 timesteps viewable in supporting media in the online journal.  Arrows indicate the chronological sequence.  Below the magnetograms is a full--disk integrated GOES X--Ray curve $(1.0-8.0\text{\AA} )$ over the course of the time series, with vertical lines showing the times of the four magnetograms.

\begin{figure}[ht]
  \capstart
  \begin{center}
    \includegraphics[width=0.5\textwidth]{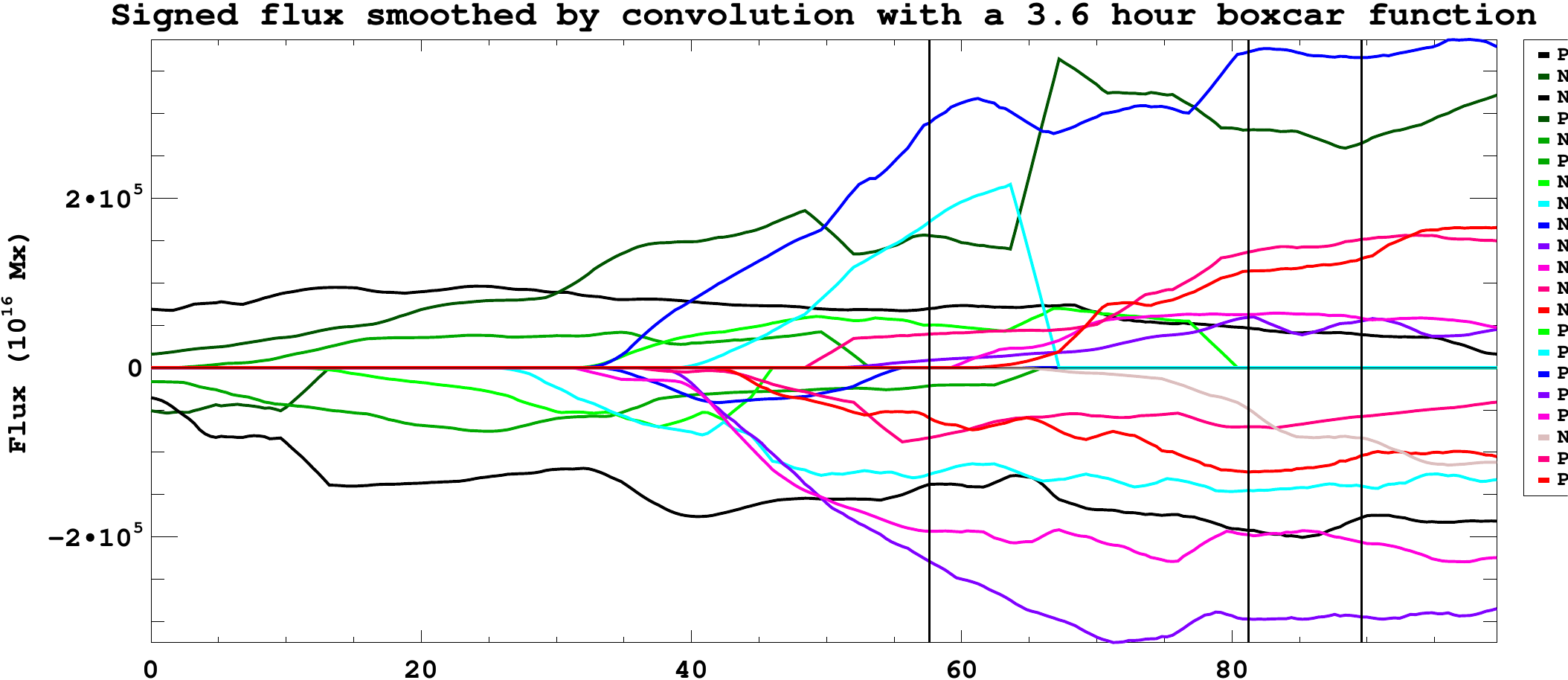}
    \caption[Smoothed regional flux]{\label{fig:sflux} Smoothed flux in each region having at least $4\times 10^{20}\unit{Mx}$ at some time.}
  \end{center}
\end{figure}

In preparation for the transition to an MCT--MCC analysis of the system, we represent each distinct region as a magnetic point source, or \emph{pole}, in the local tangent plane at each timestep.  For consistency from timestep to timestep, our point of tangency at each time is taken as the center of charge at the initial timestep, migrated through solar rotation to the present timestep.  Pole $j$ at time $i$ is defined by its associated region's total flux $\psi^i_j$ and flux--weighted centroid $\bar{\vect{x}}_j^i$:
\begin{gather}
  \label{eq:pole}\psi^i_j = \int_{\mathcal{R}^i_j} B_z(x,y)\, dx\, dy \\
  \label{eq:centroid}\bar{\vect{x}}_j^i = (\psi_j^i)^{-1}\int_{\mathcal{R}_j^i} \vect{x} B_z(x,y)\, dx\, dy \quad.
\end{gather}
The vertical magnetic field $B_z(x,y)$ accounts for both line--of--sight effects, by assuming a radial field, and pixel foreshortening, as described above.  To reduce the effects of noise in our masking algorithms, we additionally smooth the flux in each region over the four day series by convolution with a 9 timestep (3.2 hour) boxcar function, using edge truncation.  For instance, pole $j$'s flux at time $i=2$ is averaged to $\bar{\psi}_j^2=(3\times\psi_j^0+\sum_{i=1}^6\psi_j^i)$.  The resulting smoothed fluxes are shown in \figref{fig:sflux} for regions that have $\vert\psi\vert > 4\times 10^{20}\unit{Mx}$ at any timestep in the series.  Discontinuous jumps, for instance around 65 hrs, indicate merging, in this case P37 into P3.  This occurs when the distinction between separately emerged regions becomes ambiguous.

It is evident from viewing the full animation associated with magnetograms in \figref{fig:partition} that the active region complex emerges in several distinct episodes.  The complex has two distinct primary locations of flux emergence: one to the North, which leads another to the South by about $35\unit{Mm}$.  The first emergence episode is ongoing at the beginning of our timeseries, at which point several smaller emerged regions are still consolidating (eg., N3/N5 in the South; N1/N2/N6, and P1/P6 in the North).  The second episode begins around our 60th timestep (30 hrs from t=0), on Feb.~12th at 16:00UT, and continues very steadily until timestep 80 (t=60 hrs, Feb.~13th, 20:00).  At this time, the Northern emergence ceases, while it appears that the Southern emergence continues, possibly in repeated $(\approx 6 hr)$ bursts.  These bursts may be an artifact of the daily variations in HMI's reported flux, noted above.

\begin{figure}[ht]
  \capstart
  \begin{center}
    \includegraphics[width=0.5\textwidth]{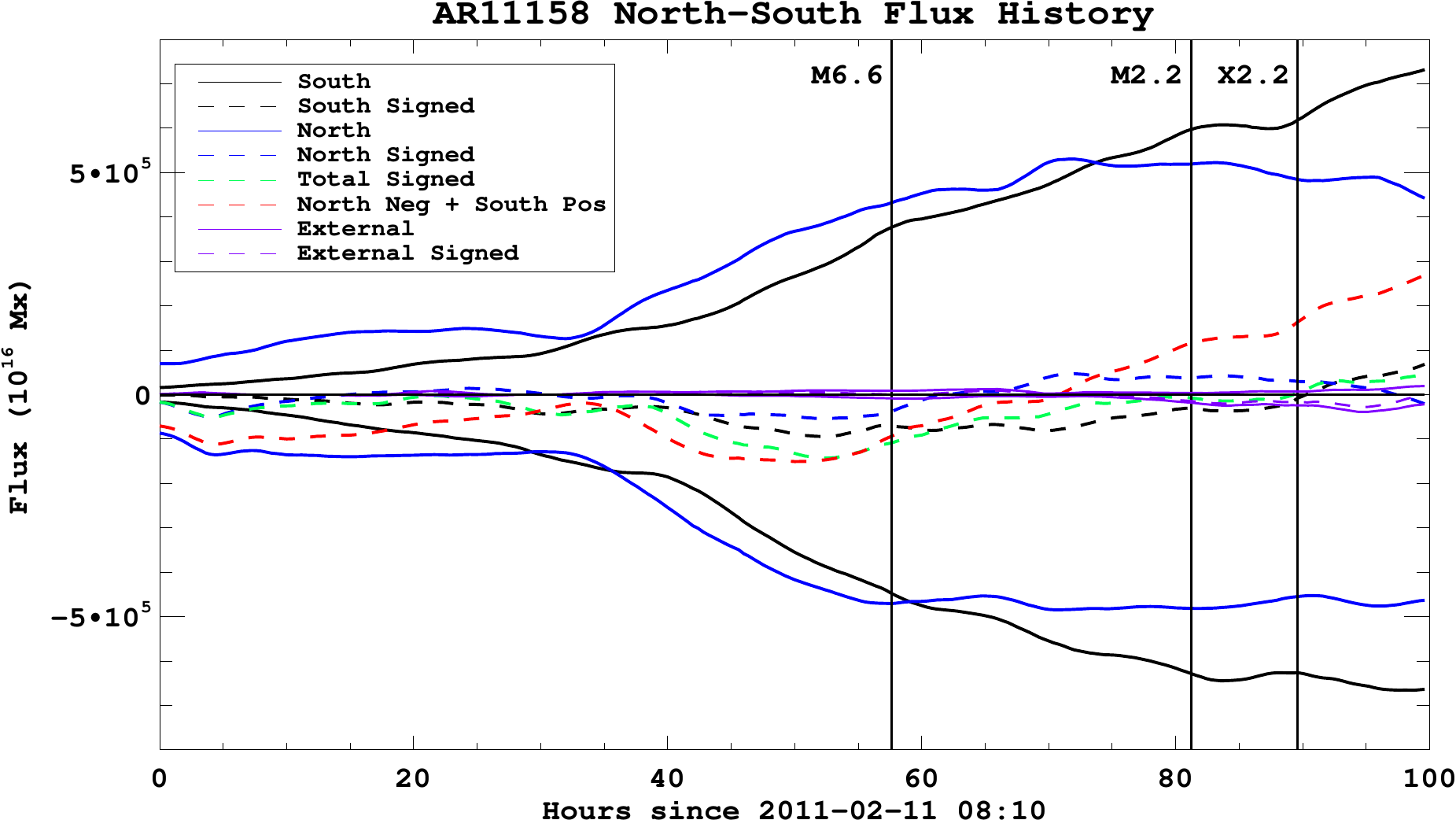}
    \caption[North--South Emergence Patterns]{\label{fig:f-imbal} Signed flux in the Northern (solid blue) and Southern (solid black) emergence zones. Dashed lines show total signed flux in North (black), South (blue), and all (red) regions, as well as the combination of Northern negative with Southern positive regions.  This readily shows the Northern and Southern emergence regions undergo two distinct emergence patterns.}
  \end{center}
\end{figure}

\figref{fig:f-imbal} shows the evolution of signed flux in those regions belonging to the Northern emergence (blue), Southern emergence (black), and a set of surrounding quiet sun (network) regions that drifted above our thresholds at various times (purple).  Of the 118 regions tracked over the timeseries, 54 belonged to the North (36 positive, 18 negative), 37 the South (13 positive, 24 negative), and 27 external to the active region complex (13 positive, 14 negative).  Dashed lines in the figure show the flux imbalance for each of the three sets, and also the flux imbalance of the total system (green).  Note that qualitatively, the Northern regions undergo a very different emergence evolution compared to the Southern regions.  This is easily seen with the dashed red line, which shows the total signed flux of the complex's central region: Northern negative flux summed with Southern positive flux.

The system's connectivity is defined by the amount of flux connecting each pole to every other pole \citep{Longcope:2009}.  This constitutes a \emph{graph}, where each pole is a vertex and each domain an edge, with the weight of each edge given by the domain flux.  The total flux of a single pole is the summed weight of all edges connected to it, and the total flux of the system of poles is the summed weight of all edges in the graph.  If there is an overall imbalance of flux, the remainder must be connected to a source located (formally) at infinity.  In general, a pair of vertices may be connected by more than one edge, and are then called multiply connected.  We have found several such instances of multiply connected vertices in AR11158, though we show below that, in this case, their effect on the system's energetics is negligible.

In determining the distribution of flux emergence within the active region, we use the method of \citet{Tarr:2012} to define a connectivity graph for the flux difference between consecutive timesteps.  This change must come in pairs, as positive and negative poles emerge and submerge together.  The total flux change between times $i$ and $i+1$ for a single pole $j$ is given by Eq.~(4) of \citet{Tarr:2012}
\begin{align}
  \label{eq:update-pol}\psi^{i+1}_j - \psi^i_j = \sum_{b}\Delta^i\Matrix{M}_{j,b} + \sum_k\Delta^i\Matrix{S}_{j,k} \quad ,
\end{align}
where $\Delta^i\Matrix{M}_{j,b}$ describes any shift in the boundary between like--signed region $b$ adjacent to $j$ and $\Delta^i\Matrix{S}_{j,k}$ describes any change in the photospheric field itself.  The former is a graph with edges connecting like signed regions of opposite flux change (flux that one region loses, another gains), while the latter is a graph connecting opposite signed regions with same sensed flux change.  The algorithms for determining these are fully described in \citet{Tarr:2012}.  To accurately deal with the two regions of emergence, we first calculate the matrix $\Delta^i\Matrix{S}$ for Northern and Southern regions separately, then combine the two resulting connectivity graphs.  Finally, we allow for connections between North and South using any remaining flux change.

We may quantify our success at capturing the flux--change processes by reconstructing the total flux of a region using its initial flux and elements of the surface change matrix.  At time $i$, we estimate a region $j$'s flux as $\psi_{j}^i = \psi_j^0 + \sum_{l=0}^{i-1} \sum_k\Delta^l\Matrix{S}_{j,k}$.  Summing these reconstructed fluxes over a set of like--signed regions connected by internal boundaries, say all the positive flux in the Southern emergence zone, should represent the total emergence of the collected regions.  We find that our method always underestimates this emergence.  Over the entire time series, we find a maximum discrepancy of between $8\%$ and $25\%$, depending on the group we reconstruct (Northern positive, Northern negative, Southern positive, Southern negative).  We believe this conservative attribution of flux change to emergence or submergence processes stems from a greedy boundary--change algorithm, asymmetries in the concentration of newly emerged positive and negative flux, and the diurnal variations due spacecraft motion, noted above.  This forces the attribution of $8-25\%$ of flux change to emergence (or submergence) with sources formally at infinity.  

Finally, we note that there is quite a bit of variation in our underestimate of emerging flux.  Our algorithm has the greatest underestimate when reconstructing the Northern positive flux emergence regions.  While at one point it is only able to pair up $75\%$ of the actual flux change, it spends fully half of all timesteps able to pair at least $85\%$ of the flux change.  The flux change formally paired with sources at infinity generally rises over the timeseries, and peaks at $13.4\%$ of the total instantaneous unsigned flux 12 hours before the M6.6 flare, then varies between $11-13.25\%$ for the rest of the series, ending 10 hours after the X2.2 flare.  This variation is consistent with the $2.7\%$ daily variation in unsigned flux found by \citet{Liu:2012}. The flux change assigned to sources at infinity at the times of the M6.6, M2.2, and X2.2 flares are $11.4\%$, $12.2\%$, and $13.3\%$ of the instantaneous unsigned flux, respectively.

\subsection{\label{sec:mod-cor}Modeling the Coronal Field}
Having quantified the connectivity graph for flux change, we may use the topological methods of \citet[\S5]{Tarr:2012} to calculate the free energy stored in coronal fields.  At every timestep we determine the potential field connectivity matrix, $\Matrix{P}^i$, using the Monte Carlo method of \citep{Barnes:2005}.  At the timesteps immediately preceding each M and X class flare, we calculate the system's potential field topology in terms of poles, nulls, and separators \citep{Longcope:2002}.  In our analysis, deviations from a potential field configuration take the form of differences in the amount of flux (either more or less) in the real field's domains versus the potential field's domains.  In the MCC model, the departure of a domain from a potential field configuration gives rise to currents in associated separators.  Every domain that is topologically linked by a separator contributes to that separator's current.  To determine which domains each separator links we use the Gauss Linking Number method of \citet{Tarr:2012}.  Completing the free energy estimate for each separator, we use the direct connection between currents flowing along separators and free magnetic energy given by \citet{Longcope:2004}.

As shown in \citet{Tarr:2012}, the self--flux of a separator (current ribbon), generated by currents flowing along it, is equal to the difference between the linked--domain fluxes in the real and potential fields:
\begin{align}
  \label{eq:fce}\psi_\sigma^{(cr)i} &= \psi_\sigma^i - \psi_\sigma^{(v)i} = \sum_D\Matrix{F}^i_D - \sum_D\Matrix{P}^i_D\quad ,\\
  \label{eq:totfcr} &\equiv - \sum_D\sum_{j=0}^{i-1}\Delta^j\Matrix{R}_D\quad .
\end{align}
Here, $\Matrix{F}^i_D$ is the real domain flux of a domain $D$ at time $i$, given by the initial potential field flux and the summation over time of the surface flux change matrix, defined above: 

\begin{equation}
  \Matrix{F}^i_D = \Matrix{P}^0 + \sum_{j=0}^{i-1}\Delta^j\Matrix{S}_D \qquad .
\end{equation}
The difference between the real and potential field fluxes at each time gives $\Delta^j\Matrix{R}_D$, the total amount of flux which may be redistributed between domains in a reconnection event.  The sum of $\Delta^j\Matrix{R}_D$ over all times up to $i$, and over all linked domains $D$, results in Equation \eqref{eq:totfcr}.

We may extend this model to include reconnection by considering the effect of reconnection at some time $k$ on the connectivity matrices described above.  The potential connectivities $\Matrix{P}^k$ do not change, as they are uniquely determined by the photospheric boundary at any time.  The only effect is to transfer flux between domains in the real field.  We accomplish this by adding some flux transfer matrix $\Matrix{X}^k$ to $\Matrix{F}$ at time $k$, so that
\begin{align}
  \label{eq:xmat} \Matrix{F}^k_{\text{postflare}} = (\Matrix{F}^k_{\text{preflare}}+\Matrix{X}^k).
\end{align}
The total flux through the photosphere does not change during a flare, so the potential field $\Matrix{P}^k$ also does not change.  According to Equations \ref{eq:fce} and \ref{eq:totfcr}, then, there is an opposite assignment of flux in the redistribution matrix $\Matrix{R}^k_{\text{postflare}}= \Matrix{R}^k_{\text{preflare}}-\Matrix{X}^k$.  We may model the effect of multiple reconnections by adding/subtracting flux transfer matrices $\Matrix{X}^l,\ \Matrix{X}^m,\ \Matrix{X}^n,\ldots$ as necessary.  Therefore, the separator self--flux at any time $i$, including all past reconnection events, is given by 
\begin{align}
  \psi&_\sigma^{(cr)i} = - \sum_D\sum_{j=0}^{i-1}\Bigl(\Delta^j\Matrix{R}_D + H(j-k)\Matrix{X}_D^k \notag\\
  \label{eq:fcr}& + H(j-l)\Matrix{X}_D^l +H(j-m)\Matrix{X}_D^m \ldots \Bigr) \ ,
\end{align}
where $H(j)= \{0, j<0; 1, j>0\}$ is the Heaviside step function.  In the following section, we will propose a minimization scheme for estimating the reconnection matrix $\Matrix{X}$ at the time of a flare.

Having thus determined each separator's self--flux at any time, we follow \citet{Longcope:2004} to relate that self--flux \eqref{eq:fce} to the separator current by 
\begin{gather}
  \label{eq:crpsi}
  \psi^{(cr)i}_\sigma = \Frac{I L}{4 \pi}\ln\Biggl(\Frac{e I^*}{\vert I\vert}\Biggr)\quad .
  \intertext{The fuctional inversion}
  \label{eq:cr}
  I(\psi^{(cr)i}_\sigma) = I^*\Lambda^{-1}(4\pi \psi^{(cr)i}_\sigma/LI^*)
\end{gather}
with $\Lambda(x) = x\ln(e/\vert x\vert) $ allows us to represent the current in terms of the fluxes.  Finally, from \citet{Longcope:2004} equation (4) we can calculate the energy in the MCC model in excess of the potential field magnetic energy,
\begin{equation}
  \label{eq:wmcc}
  \Delta W_{MCC} = \Frac{1}{4\pi}\int_{\Psi_{potl}}^{\Psi}I d\Psi = \Frac{L I^2}{32 \pi^2}\ln\Bigl(\Frac{\sqrt{e} I^*}{\vert I\vert}\Bigr)
\end{equation}
which, via equation \eqref{eq:cr}, is a function of the calculated separator fluxes $\psi^{(cr)i}_\sigma$.

We determine the coronal topology for the M6.6 flare at 17:22 on Feb.~13, 6 minutes before flare onset, and 16 minutes before GOES peak intensity.  At this time, our model consists of 27 sources (16 negative, 12 positive) and 26 nulls.  Following \citet{Longcope:2002}, these numbers satisfy both the 2D and 3D Euler characteristics, so we believe we have found all nulls.  Every null is prone, and there are no coronal nulls.  One null is asymptotic in the sense of \citet{Longcope:2009}, laying along the direction of the region's dipole moment computed about the center of unsigned flux, $\mu$, at a distance $r_0 = 2\mu/q_\infty$, where $q_\infty$ is the net charge.  This null's separatrix surface forms a boundary between the region's closed flux and surrounding open flux.  7 additional source--null pairs are part of unbroken fans: P57/B22, P52/B21, N51/A20, N45/A18, N43/A16, N42/A25, and N38/A26.  Using these values and the equation between (26) and (27) of \citet{Longcope:2002}, we expect to find 17 separators in the corona (along with 17 mirror separators), which we do find.  We therefore believe we have completely specified the system's topology on the eve of the flare.  

We perform a similar analysis just prior to the M2.2 and X2.2 flares.  While we do not find every topological element in these later flares, we believe we find all that play a significant role in each case.

For the 17 separators at the time of the M6.6 flare, we find 90 linked domains.  Two of the separators have the same endpoints, nulls A07/B01, and therefore enclose multiply connected source pairs \citep{Parnell:2007}.  These are known as redundant separators.  In this case, the two separators lay nearly along the same path, implying a slight wrinkle in the intersecting fan surfaces.  This creates one additional flux domain enclosed by the two separators.  Because the cross--sectional area in this case is small, the enclosed flux is small relative to the total flux enclosed by each separator, and the corresponding energy due to the redundant separator is negligible.  The Monte Carlo estimate of fluxes enclosed by the different separators used 500 field lines.  There was no difference in the number linked by the separators, so we conclude that the fluxes they link are identical to a fraction of a percent.  We use only one of the two in our calculation, and arrive at the same result independent of this choice.

\section{\label{sec:energy}Energy estimates}
As stated above, one shortcoming of current MCT/MCC analyses are their inability to account for violation of the flux constrains.  As such, they have no way to account for reconnection.  We here present a method for relaxing those constraints at any timestep, while allowing the system to continue evolving after reconnection.

During reconnection, flux is exchanged across the field's separators.  Each separator lies at the boundary of four flux domains, and the separators involved in the flare identify the set of domains which exchange flux.  Some of these domains are flux superficient, containing more flux than in a potential field, and some deficient.  Not every separator needs to be involved in every flare, and not all flux is necessarily transferred in every flare, even within the subset of involved domains.  

Reconnection does not simply involve the transfer of flux from superficient to deficient domains.  Two domains on opposite sides of a separator\footnote{A separator connects two null points of opposite sign, each of whose two spines connect to sources of the same polarity.  The four domains form all possible connections between the two positive and negative spine sources.  We designate two domains ``opposite'' if they share no spine sources.} (X--point in 2D) reconnect fieldlines, transferring flux to the remaining two domains.  There is no physical reason why opposing domains must both have more (or less) flux than a potential field configuration.  Instead, the state of the current domain depends on the history of its poles: where and with whom they emerged, who they reconnected with in the past, and what their current geometric orientation is.  The only requirement for reconnection is that the two flux--donating domains contain some flux (nonzero elements of $\Matrix{F}$ in Equation \eqref{eq:xmat}).  In such cases, this poses the interesting question of whether the potential field configuration is always attainable through reconnection, or if there exists some local minima in configuration space.  We briefly consider this below, but leave a more detailed analysis of the question for another investigation.

We use a simple iterative minimization algorithm to model the redistribution of flux across a set of separators involved in a particular event.  For each separator, we propose a small transfer of flux between the four surrounding domains.  Each domain is specified by the indices of its two poles in a connectivity matrix, and reconnection is described by the transfer of flux $d\psi$ from two elements of our redistribution matrix to two others.  Supposing that the two flux--donating domains are $\{j,k\}$ and $\{l,m\}$, while the two receiving domains are $\{n,o\}$ and $\{p,q\}$, then the 4 involved elements of redistribution matrix change as follows: $\Delta^i\Matrix{R}_{jk/lm}\rightarrow -d\psi \text{ and } \Delta^i\Matrix{R}_{no/pq}\rightarrow +d\psi$.  We then calculate the total energy drop over the entire system (the summation of Equation \eqref{eq:wmcc} for every separator).  We apply the small reconnection resulting in the largest total free energy drop.  We then calculate the next set of energy drops for each small reconnection, and this process continues until one of two conditions is met: either 1) no flux transfer is possible because no pair of opposing domains both have nonzero elements of the connectivity matrix $\Matrix{F}$; or 2) any small reconnection increases the total free energy.  The sum of every small reconnection $d\psi$ then fills out the flux--transfer matrix $\Matrix{X}^k$ for the flare at time $k$.

\section{\label{sec:ribbons}Observations of flare ribbons}

Having determined how to model reconnection, we now turn our attention to determining when to apply a minimization.  At present we have no model for the mechanism in the actual corona which initiates reconnection at a current sheet.  All we can infer, from observations of actual flares, is that at some instant the reconnection does begin at certain separators.  We therefore rely on observations of this kind to determine the separators undergoing reconnection, and when this reconnection occurs.  

For the present study, we perform a minimization for every flare associated with AR11158 with GOES class of M1.0 or greater.  This pares the number of separate minimizations down to a manageable amount for a proof of concept, while still allowing us to understand some of the large scale processes at work in the active region's evolution.

We employ chromospheric data to select a subset of coronal domains involved in each flare.  We associate the chromospheric flare ribbons observed in the 1600\AA\ channel of AIA data with specific spine field lines of the topological skeleton.  While the spine lines do not always geometrically match the ribbons, there is a topological correspondence \citet{Kazachenko:2012}.  The spines are the photospheric footpoints of reconnecting loops, and therefore indicate which flux domains are involved in each flare.  Magnetic reconnection across a separator couples the flux redistribution in the corona to the photospheric spines of the separator's null points.  The highlighting of spine lines by flare ribbons thus indicates those separators involved in each flare.  To make use of this information, we relax flux constraints (allow for reconnection between four domains) using only those separators associated with the highlighted ribbons.

\begin{figure}[ht]
  \capstart
  \begin{center}
    \includegraphics[width=0.8\textwidth]{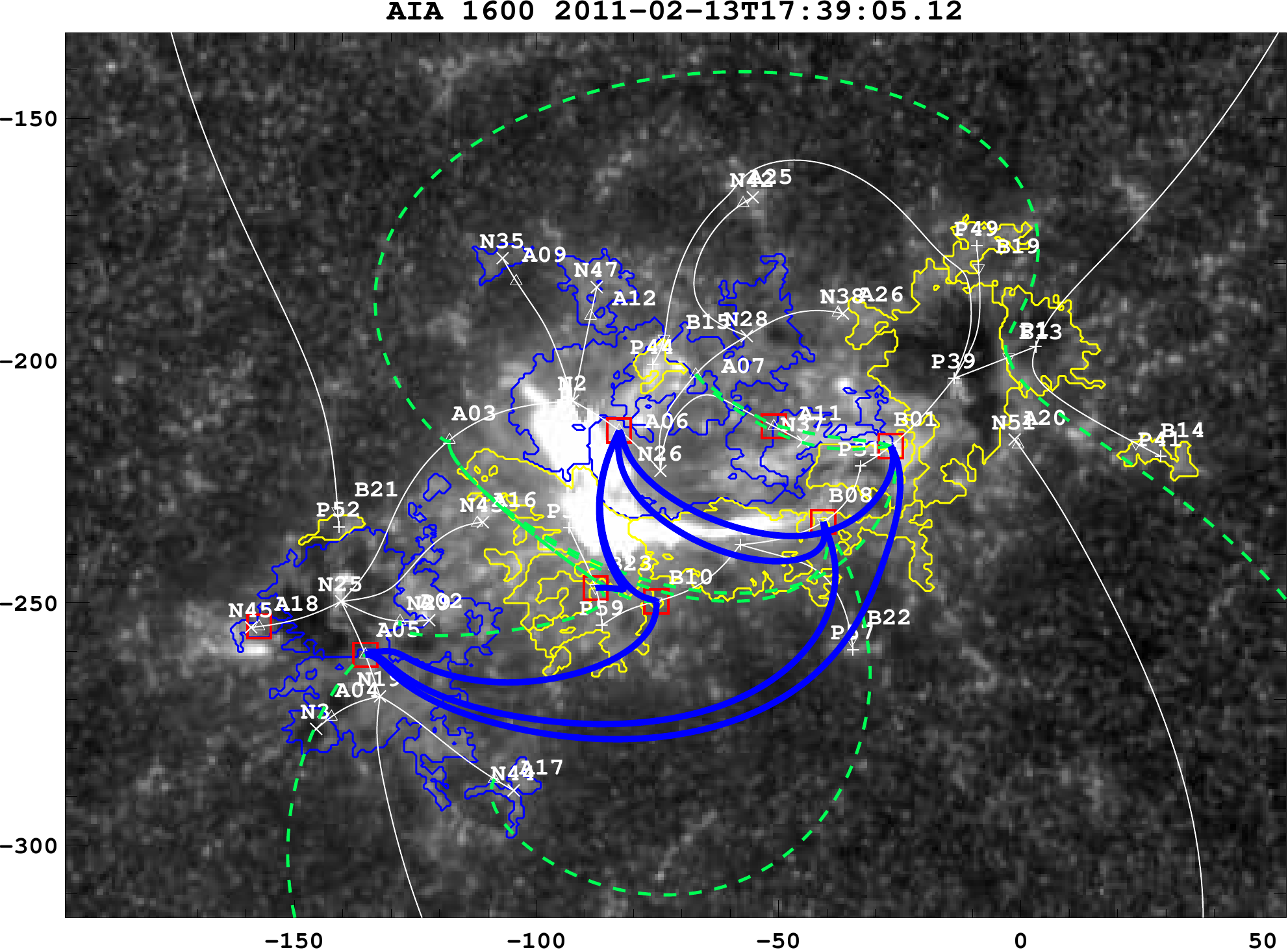}
    \caption[M6.6 Flare Ribbon]{\label{fig:143ribbon} Log--scaled AIA 1600\AA\ image during the GOES class M6.6 flare, with coordinates given in arcseconds from disk center.  The grayscale saturates at 6000 counts, roughly half the pre--flare maximum pixel value.  The potential field skeleton is overlaid: pluses and crosses are positive and negative poles, respectively; triangles are positive ($\vartriangle$) and negative ($\triangledown$) nulls; thin solid white lines depict spines.  The energy calculation only attempts reconnection across those separators having two boxed nullpoints as footpoints.  These seven separators displayed as thick solid blue lines.  The remaining ten separators are displayed as green dashed lines.}
  \end{center}
\end{figure}

Figure \ref{fig:143ribbon} shows the AIA 1600\AA\ data for a selected timestep during the M6.6 flare.  The AIA image is displayed in a logarithmic grayscale, and shows a relatively simple two ribbon flare.  Overplotted are contours of the magnetogram at +75G (yellow) and -75G (blue), as well as the topological skeleton (see caption for details, and online material for an animation covering the time of the flare).  Clearly visible are the two primary flare ribbons, located on either side of the central polarity inversion line (PIL between Southern--emerged positive flux and Northern negative flux).  Evident in the online animation are several other, smaller flare ribbons, located in the Southern negative and Northern positive regions.

There are spine fieldlines associated with each ribbon.  The Northern primary ribbon corresponds to the spine lines of null A06, between poles N2 and N26, near $(-90,-210)^{\prime\prime}$.  The Southern ribbon is only morphologically similar the potential field MCT model.  We separate the more diffuse P59 region from the more concentrated P3 and P37, which forces the creation of two null points (B23, B10) with associated spine lines, near $([-90/-75],-250)^{\prime\prime}$, respectively.  We believe the real field likely has a null directly between P3 and P37, creating a more direct spine line between the two.  

The spines involved in this flare have the red--boxed nullpoints in Figure \ref{fig:143ribbon} as their spine sources.  This indicates that flux is transferred only across those separators connecting two of the boxed nulls.  The projection onto the photosphere of seven such separators in this flare are shown as thick blue lines.  The remaining separators are shown as dashed green lines.  The free energy of these other separators may still change during the flare despite have no reconnection across them, provided the Gauss Linking Number between the separator field line and any domain which does participate in reconnection is nonzero, as indicated by Equation \eqref{eq:fcr}.

\begin{figure}[ht]
  \capstart
  \begin{center}
    \includegraphics[width=0.5\textwidth]{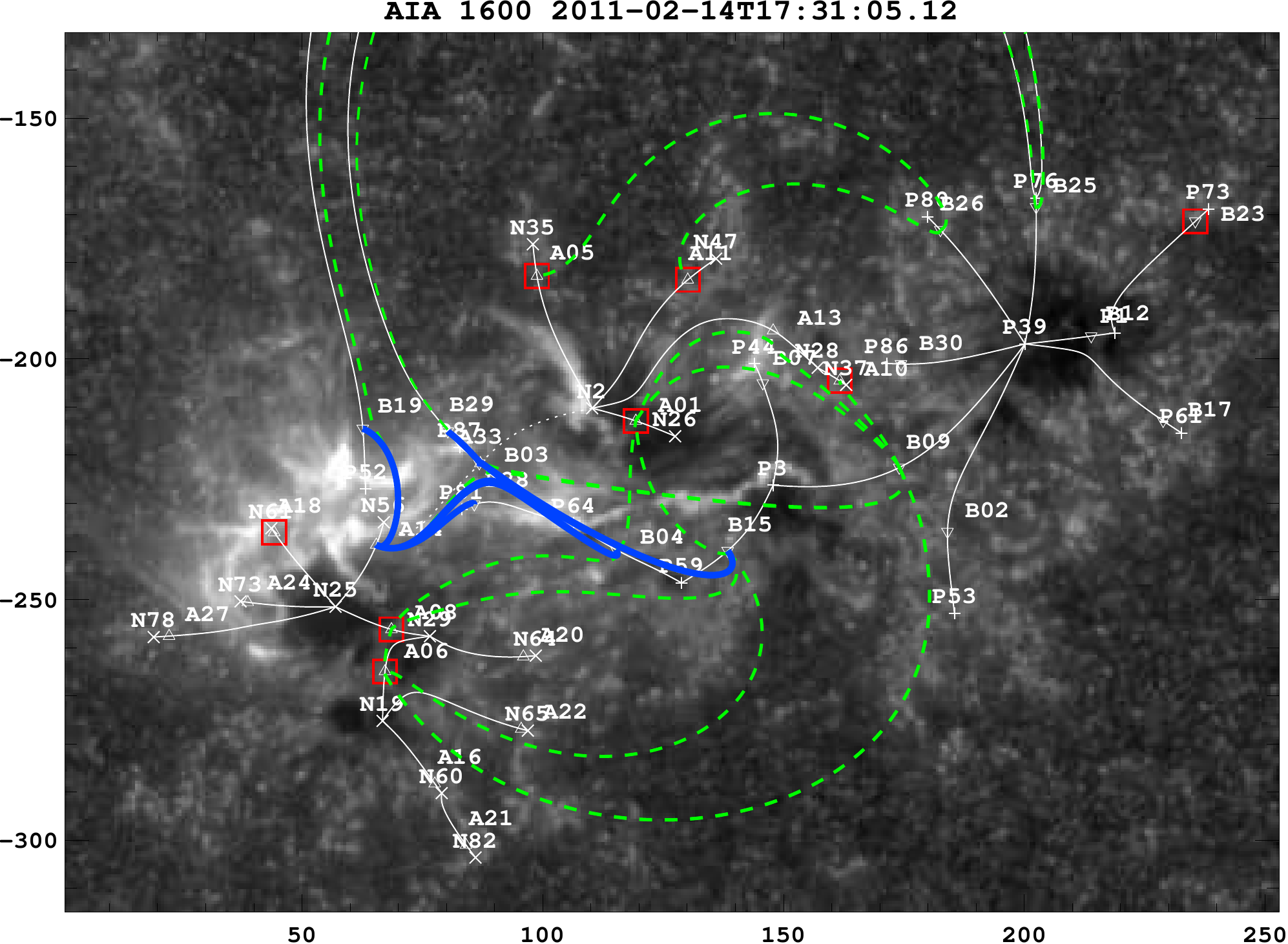}
    \caption[M2.2 Flare Ribbon]{\label{fig:202ribbon} Same as \figref{fig:143ribbon}, for the M2.2 flare.  Solid blue and green lines show separators connected to nulls with spines laying approximately along paths of flare ribbons observed in AIA 1600\AA\ channel.  Locations of other separators are shown as dashed green lines.}
  \end{center}
\end{figure}

\begin{figure}[ht]
  \capstart
  \begin{center}
    \includegraphics[width=0.5\textwidth]{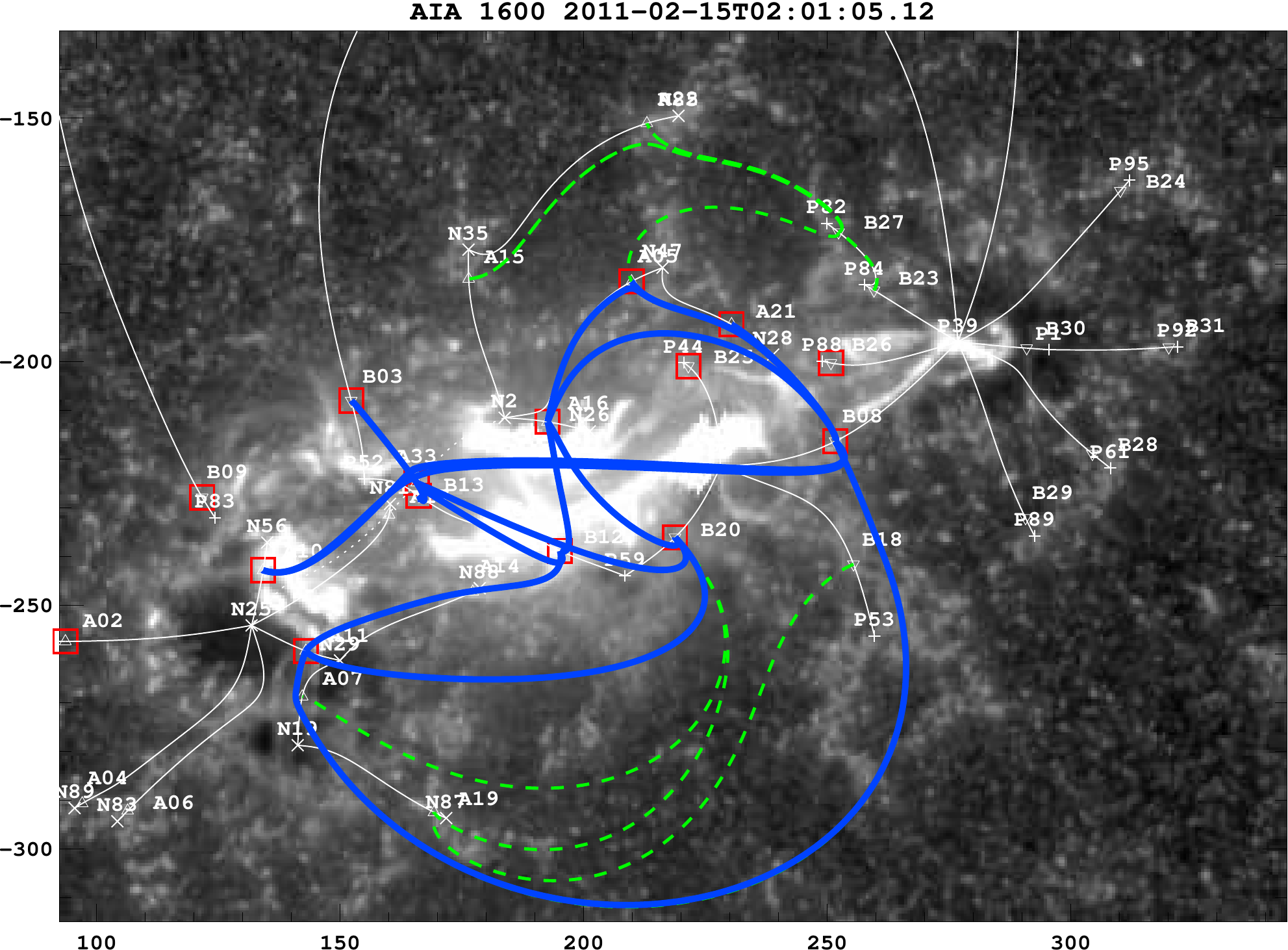}
    \caption[X2.2 Flare Ribbon]{\label{fig:224ribbon} Same as \figref{fig:143ribbon}, for the X2.2 flare.}
  \end{center}
\end{figure}

Figures \ref{fig:202ribbon} and \ref{fig:224ribbon} are similar to Figure \ref{fig:143ribbon}, during the M2.2 and X2.2 flares, respectively.  We have left off the contours of the magnetogram in these figures for clarity.  The X class flare in particular is more complex than previous flares, involving more and disparate parts of the active region complex.  This increased activity is likely influenced by the creation of a coronal null point just prior to the M2.2 flare, whose fan surface effectively separates the Northern and Southern emergence zones.

The energy buildup prior to the M6.6 flare is particularly dependent on the emergence of N26 in the North.  This generates the null point in the North to which the four Northern involved separators attach.  In the 25 hours between N26's emergence at 2011-02-12 16:00 UT and the M6.6 flare, we calculate an increase in free energy due to currents along these four separators of $2.87 \times 10^{31}\unit{erg}$, about one third of the total MCC free energy in the system at this time.  These separators link domains N26/P37, N26/P59, N26/P31, N26/P39, N26/P44, N28/P31, N37/P3, and N37/P39.

The 1600\AA\ flare ribbons indicate that 6 separators are involved in the M2.2 flare (\figref{fig:202ribbon}).  Four connect to null A14 between regions N25 and N56, and 2 connect through the coronal null A33; their projections in the photospheric plane are shown as solid blue solids, with the remaining separators shown as dashed green lines.

\begin{figure*}[ht]
  \capstart
  \begin{center}
    \includegraphics[width=0.8\textwidth]{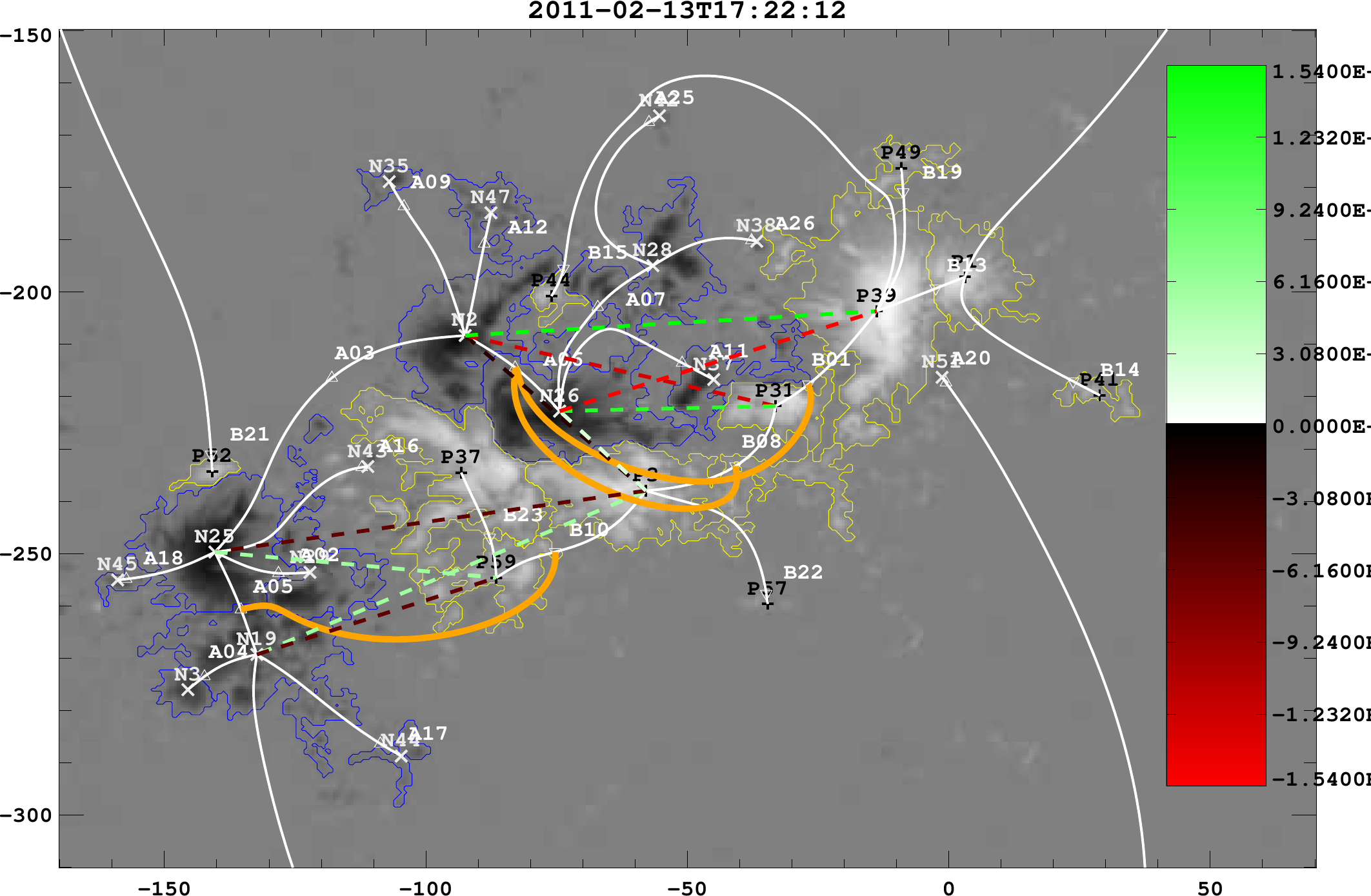}
    \caption[M6.6 Minimization]{\label{fig:m66min} Flux redistributed as a result of energy minimization.  The background image shows the magnetogram, mask, poles, nulls, and spine field lines.  Thick orange lines show the three separators utilized during the minimization.  Dashed lines indicate domains involved in the minimization, with colors representing the amount of flux gained (white to green) or donated (black to red).  The colorbar scale is in units of $10^{16}\unit{Mx}$.}
  \end{center}
\end{figure*}
\figref{fig:m66min} illustrates the result of the free energy minimization for the M6.6 flare.  Thick orange lines show the projections of separators across which flux was transferred by the minimization process.  Note that this involves three separators of the six identified via flare ribbons in \figref{fig:143ribbon}, which are themselves a subset of the 17 total separators at this time.  The dashed lines overlaid on the magnetogram, mask, and topological footprint background show those domains that exchanged flux during minimization---these are not fieldlines, just identifications of the involved domains.  The amount of flux loss or gain is indicated by the colorbar, with white to green indicating increasing amounts of flux gain, and black to red increasing flux donation.  

It is immediately apparent that the minimization does not exactly match our expectation from the flare ribbons, despite our specification of ``involved separators.''  In particular, the four domains involving N2, N26, P37, and P59 are essentially nonparticipants in modeled flare, whereas they are clearly the dominant players in the actual flare.  In our model of emergence, these domains are not simply flux deficient relative to the potential field, but have \emph{zero} initial flux.  We will discuss this in more detail in \S\ref{sec:disc}.

In total for the M6.6 flare minimization, we find that $4.2\times 10^{20}\unit{Mx}$ of flux was exchanged between $10$ domains across $3$ separators.  This exchange took $1922$ iterative minimization steps, resulting in a total drop of $E_{\unit{drop}}=3.9\times10^{30}\unit{erg}$, $2.5\%$ of the pre--minimization MCC free energy ($E_{\unit{MCC}}=1.5\times 10^{32}\unit{erg}$) and $1.1\%$ of the potential field energy ($E_{\unit{potl}}=3.9\times 10^{32}$).  These results are summarized in Table\ref{tab:minimize}, together with those for the M2.2 and X2.2 flares, and we discuss them in more detail in the next section.

 \begin{table*}[ht]
   \centering
     \caption{\label{tab:minimize}Summary of energy minimization for each flare.  Columns are 1) GOES class; 2) Flux exchanged by the minimization algorithm; 3) Number of domains involved in the minimization; 4) Number of separators across which flux is exchanged; 5) Number algorithm steps; 6) Initial free energy of the MCC; 7) Energy drop due to minimization; and 8) Potential energy using the magnetogram as a lower boundary.}
   \begin{tabular}{lccccccc}
     \tableline\tableline
     Flare & Flux ($10^{20}\unit{Mx}$) & Domains & Separators & Steps & Initial $E_{\unit{MCC}}$ & $\Delta E_{\unit{MCC}}$ &  $E_{\unit{potl}}$ \\
     \tableline 
     M6.6 & 4.2  & 10 & 3  & 1922  & $1.53\times 10^{32}$ & $3.89\times 10^{30}$ & $3.83\times 10^{32}$ \\
     M2.2 & 2.0  & 15 & 8  & 327   & $1.65\times 10^{32}$ & $2.62\times 10^{30}$ & $5.77\times 10^{32}$ \\
     X2.2 & 21.0 & 17 & 10 & 29504 & $2.94\times 10^{32}$ & $1.68\times 10^{32}$ & $5.55\times 10^{32}$ \\
     \tableline
   \end{tabular} 
 \end{table*}

\section{\label{sec:disc}Discussion}

This investigation builds on \citet{Tarr:2012} in adding the observational history of an active region's flux evolution, in particular its flux emergence, to the MCC model.  Here we have relied on line--of--sight magnetograms provided by SDO/HMI and generated our flux histories by assuming a radial field.  With the arrival of the HMI Active Region Patches (HARPs) dataseries to JSOC, future investigations can use the actual vertical flux determined by HMI's vector magnetograms.  

We were fortunate in the present case to have HMI observe the entire history of AR11158 from its emergence around $50^\circ$ Solar East on Feb.~10th, 2011, to its rotation off-disc some 9 days later.  In the more common case where we do not observe the entire emergence of an active region, we could use a NLFFF extrapolation to generate an initial connectivity state, which would then be updated in time via the methods of \S\ref{sec:mod-phot}.

We have gone several steps further than previous energy estimates using the MCT/MCC framework, such as \citet{Tarr:2012} or \citet{Kazachenko:2012}.  Most importantly, we now allow for violation of the no--reconnection flux constraints of the MCC.  This enables us to not only consider energy storage due to currents along a subset of separators in a flare, but also allows for an estimate of flux transfer and energy conversion during a flare.  Further work may yield an interesting comparison between the flux transferred in our minimization to estimates of reconnected flux based on analysis of flare ribbons, as was done in \citet{Longcope:2007}.

For each flare, we choose a subset of separators that are allowed to transfer flux during our minimization based on observations of chromospheric flare ribbons.  Each separator bounds 4 domains, though some of these domains are bounded by more than one separator.  Thus, the number of domains which may exchange flux via reconnection is typically less than 4 times the number of involved separators.  

It is interesting that the minimization scheme we have proposed does not utilize every allowed domain.  As mentioned above, in order to undergo reconnection, two domains on opposite sides of a separator must both contain flux (have nonzero elements in the connectivity matrix $\Matrix{F}$).  Focusing again on the M6.6 flare, the two separators that connect to null A06, between poles N2 and N26, and have pole P59 as a spine source of their B--type nulls (nulls B10 and B23) contain very strong currents and border flux domains with highly nonpotential connectivities.  And yet, they do not participate in the minimization.  In this case, there is no pathway of free energy loss that results in donatable flux on opposite sides of these separators.  The lack of any flux in the P59/(N2, N26, N28) domains effectively cuts off any involvement of P37 in our model of this flare.

Between poles N2, N25, P37, and P59, flux transfer may occur in either of two directions.  Domains N2/P59 and N26/P37 may donate flux, with N2/P37 and N26/P59 receiving, or N2/P37 and N26/P59 may donate flux with N2/P59 and N26/P37 receiving.  In either case, one of P59's domains must donate flux.  Since both the P59/N2 and P59/N26 domains have \emph{no} flux in our model (not to be confused with having less than the potential field configuration), this reconnection cannot occur.

We do, however, observe the primary flare ribbons involving just these domains.  So, where has our model of this active region gone awry?  First, we note that there are many small flare events smaller than M1.0 prior to the first flare we consider.  Any of these events may transfer flux into the domain necessary for the M6.6 event.  

Second, it is likely that a great deal of reconnection takes place that is unassociated with any GOES class flare.  This may be a ``steady'' reconnection, as is observable in EUV and X--ray images of emerging magnetic regions.  For instance, in our previous study \citep{Tarr:2012} of AR11112 from October 2011, we saw that newly emerged flux steadily reconnected with surrounding preexisting flux, as evidenced by the encroachment of the bright XRT kernel into the surrounding flux.  This is discussed in detail in \citet{Tarr:2013b}.  Although the region did produce several flares later in its evolution, this steady reconnection occurred independently of any observable flares.  We believe that a similar process is ongoing in AR11158's evolution.  In particular, we believe this type of steady reconnection occurs along the central PIL, where Southern emerged positive flux has collided with and sheared relative to Northern emerged negative flux.  Such a steady, low--level reconnection would populate these central domains with flux.  The free energy of this portion would increase as these domains continued to shear, culminating in the series of explosive reconnections observed at later times.  One may see evidence of this in EUV images from AIA, showing low laying loops that cross the central PIL, together with very high loops apparently connecting the most Westward positive flux to the most Eastward negative flux concentrations.

Another interesting point is that, in our minimizations, no separator expelled all of its current and thereby reached the potential field state, as one might expect to be the case in an MCC model\footnote{In other types of models \citep[e.g.,]{Regnier:2007}, the potential field state is \emph{not} generally accessible via reconnection.  For instance, in models where helicity is conserved, reconnection drives the system towards the linear force free field with the same helicity.}. Instead, in each of the three reconnection events modeled here, the system reaches a state where reconnection across on separator reduces one domain's flux to zero.  Reconnection across a second separator can repopulate the zero--flux domain, allowing further reconnection across the first separator.   However, it often happens that the two separators each require flux donation from that same domain, and so no further reconnection is possible across either one.  \citet{Longcope:2010} also found that the total free energy derived from the MCC model was greater than their calculated energy losses observed during the Feb. 24, 2004 X class flare.  They attributed this to incomplete relaxation through reconnection, but had no way to assess why this might be the case, as we have developed here.

\begin{figure}[ht]
  \capstart
  \begin{center}
    \includegraphics[width=0.5\textwidth]{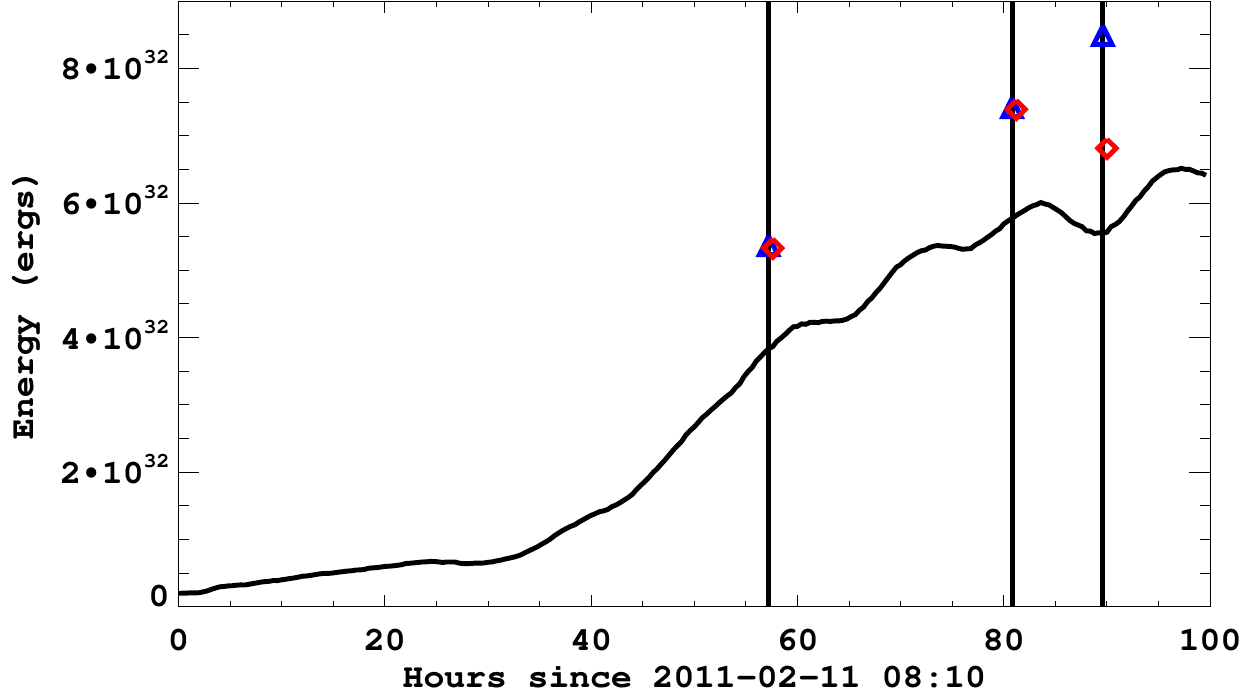}
    \caption[X2.2 Flare Ribbon]{\label{fig:Eplot} The potential energy (solid line) and MCC free energies both before (triangles) and after (diamonds) each minimization, the times of which are indicated by vertical lines.  There is no way to calculate a potential energy in the MCT/MCC model, so the free energy has been plotted above the potential energy derived from a Fourier transform method.}
  \end{center}
\end{figure}

We may compare the energies associated with each event in a variety of ways.  The solid line in \figref{fig:Eplot} shows the potential energy calculated from each magnetogram using the Fourier Transform method of \citet{Sakurai:1989}.  This is similar to, but not directly comparable with, the potential energy shown in Figure 4(c) of \citet[][(hereafter S12)]{Sun:2012}.   When not stated explicitly in the text, we have determined approximate values for potential and free energies from their Figure 4.  

Because S12 use the vertical flux determined from the HMI vector magnetograms whereas we use the LOS field, deprojected assuming a radial field at every pixel, we have different lower boundaries for determining the potential energy, so that our calculated values differ.  Our potential energy appears consistently lower than that of S12, and also seems to have more substantial variations.  These variations become more pronounced after $t=65\unit{hrs}$, when the Northern region ceases substantial emergence and the active region crosses disk center.

The free energies of the MCC model $E_{\unit{MCC}}$ are shown in \figref{fig:Eplot} as red triangles at the time of each flare and blue diamonds after minimization.  We have added the MCC values to the calculated potential field energies at each time in order to emphasize that these are energies in excess of the potential field.  However, this addition should be taken with caution, because the free and potential energies are calculated using two incompatible boundary conditions: the former using the point sources of the MCT model with an imposed FCE constraint, the later using continuously distributed magnetograms.  The potential energy of a point source is infinite, and the MCC is only able to determine the difference between energies of the FCE and potential fields.  Even given these issues, it is still a useful comparison to make.

The initial free energy of the MCC increases from flare to flare, while the calculated energy drop due to flux exchange during minimization scales with the size of each flare.  This is most easily seen in the $\Delta E_{\unit{MCC}}$ column of Table\ref{tab:minimize}.  The general trend of the amount of flux exchange, and resulting energy drop, in each flare follows our expectations given the GOES class.  An order of magnitude more flux is exchanged for the X2.2 flare compared to the M flares, yet this leads to a two order of magnitude greater free energy drop.  This highlights the important point that free energy does not scale linearly with flux--difference from potential, even for a given separator.

S12 made extensive use of HMI's vector magnetograms to present a detailed discussion of magnetic energy in AR11158, so it is useful to compare the results of that NLFFF model to ours.  Those authors extrapolate a NLFF field at a 12 second cadence using the HMI vector magnetogram as a lower boundary.  Each extrapolation is independent of the others, maintaining no memory of previous magnetic flux or connectivity.  It can therefore be difficult to ascribe any rise or drop in the free energy from one time to the next to any particular event.  They state a spectropolarimetric noise of $\approx 4\times 10^{30}\unit{ergs}$, but acknowledge that errors due to extrapolation are unknown and possibly greater.  We note that a persistent change from before to after is more likely to be a manifestation of change in the actual solar field.

For the X2.2 flare, S12 find an initial free energy of $\approx 2.5\times 10^{32}\unit{ergs}$, and a persistent free energy drop from before to after the flare of $0.34\pm 0.04\times 10^{32}\unit{ergs}$.  This may be compared to our model, which sets the initial free energy at $2.94\times 10^{32}\unit{ergs}$ and a pre--to--post minimization drop of $1.68\times 10^{32}\unit{ergs}$.  That both of these results are greater than those of S12 is a little surprising because MCC provides a lower bound on the free energy of linear force free fields evolving quasistatically \citep{Longcope:2001}.  While it is not necessary that a LFFF have greater free energy than a NLFFF derived from the same boundary, one often assumes it to be the case.  

At the same time, the MCT/MCC model does not involve an extrapolation, except to determine the topological structure of the region.  The non--potentiality of the region is determined simply by using observations emergence to fix the regions connectivity.  Our minimization scheme only determines the total amount of energy loss provided that all possible free energy minimizing reconnections take place.  Not only may the algorithm terminate in a local free energy minimum, as discussed above, there is no physical reason why all possible reconnections need to take place in a single event.

It is much more difficult to make such a comparison between these two models for the M6.6 and M2.2 flares.  From Figure 4(d) of S12, we estimate initial free energies of $1.2\times 10^{32}\unit{ergs}$ and $2.0\times 10^{32}\unit{ergs}$ for these flares, respectively, compared to our results of $1.53\times 10^{32}\unit{ergs}$ and $1.65\times 10^{32}\unit{ergs}$.  Our minimization gives pre--to--post reconnection drops of $3.92\times 10^{30}\unit{ergs}$ and $2.62\times 10^{30}\unit{ergs}$ in each case, which is the approximate level of the spectropolarimetic noise in the NLFFF extrapolations.  Indeed, again looking at Figure 4(d) of S12, neither the M6.6 or M2.2 flare appears cotemporal with a decrease in free energy, and certainly not with a persistent decrease, as is the case with the X2.2 flare.

As a final energetic comparison, we calculate the energy loss due to radiation in each flare using the GOES light curves via the method of \citet{Longcope:2010, Kazachenko:2012}.  For the M6.6, M2.2, and X2.2 flares we find radiative losses of $1.2\times 10^{30} \unit{ergs}$, $0.5\times 10^{30}\unit{ergs}$, and $4.2\times 10^{30}\unit{ergs}$, respectively.  The above mentioned papers show that other energetic losses, such as thermal conduction and enthalpy flux, tend to dominate the radiative losses during flares.  The total energetic loss is difficult to precisely quantify, but may exceed the GOES estimated radiative loss by a factor of $\approx 2--15$.  Given that, the results of our minimization for the M6.6 and M2.2 compare favorably with the GOES estimates.  However, our estimation for energy loss during the X--class flare may be greater than the observed energetic losses by an order of magnitude.  This is not surprising because the great extent of the X--class flare ribbons encompassed more separators and more domains, ultimately allowing more pathways for energy minimizing reconnection.  This allowed the minimization algorithm to exchange much more flux before halting in a local minimum.

\acknowledgments
Graham Barnes provided code for producing the potential field connectivity matrices using a Monte Carlo algorithm with Bayesian estimates, as described in \cite{Barnes:2005}.  Development of the code was supported by the Air Force Office of Scientific Research under contract FA9550-06-C-0019.  We would like to acknowledge our use of NASA's Astrophysics Data System.  This work is supported by NASA under contract SP02H3901R from Lockheed--Martin to MSU.

\end{document}